\documentclass[
]{ceurart}

\usepackage{tikz}
\DeclareRobustCommand{\TripBadge}[3]{%
	\tikz[baseline=(b.base)]\node[draw=black, line width=0.4pt, rounded corners=1.2pt,
	fill=#1, text=#2, font=\sffamily\bfseries\scriptsize, inner sep=0pt,
	minimum width=1.35em, minimum height=1.35em] (b) {#3};}
\newcommand{\PreTripBadge}{\TripBadge{black!18}{black}{P}}
\newcommand{\InTripBadge}{\TripBadge{black!37}{white}{I}}

\usepackage{tabularx}
\usepackage{acro}
\usepackage{subfig}
\DeclareAcronym{TRS}{
	short = TTRS,
	long = travel and tourism recommender system
}
\DeclareAcronym{POI}{
	short = POI,
	long = point of interest
}
\DeclareAcronym{LLM}{
	short = LLM,
	long = large language model
}
\DeclareAcronym{LBSN}{
	short = LBSN,
	long = location-based social network
}
\DeclareAcronym{GenAI}{
	short = GenAI,
	long = generative AI
}
\DeclareAcronym{RS}{
	short = RS,
	long = recommender system
}
\DeclareAcronym{nDCG}{
	short = nDCG,
	long = normalized Discounted Cumulative Gain
}
\begin{document}

\copyrightyear{2026}
\copyrightclause{Copyright for this paper by its authors.
  Use permitted under Creative Commons License Attribution 4.0
  International (CC BY 4.0).}

\conference{Preprint}

\title{Behavior-Mining, Generative Conversations, and Collaborative Advisory: the Future of Travel and Tourism Recommender Systems}
\author[1]{Alejandro Bellogín}[%
  orcid=0000-0001-6368-2510,
  email=alejandro.bellogin@uam.es]

\author[2]{Linus W.\,Dietz}[%
  orcid=0000-0001-6747-3898,
  email=linus.dietz@kcl.ac.uk]

\author[3]{Francesco Ricci}[%
  orcid=0000-0001-5931-5566,
  email=francesco.ricci@prof.senior.unibz.it]

\author[4]{Pablo Sánchez}[%
  orcid=0000-0003-1792-1706,
  email=psperez@comillas.edu]

\address[1]{Universidad Autónoma de Madrid, Madrid, Spain}
\address[2]{King's College London, London, United Kingdom}
\address[3]{Free University of Bozen-Bolzano, Bolzano, Italy}
\address[4]{Instituto de Investigación Tecnológica (IIT), Universidad Pontificia Comillas, Madrid, Spain}

\fntext[1]{All authors contributed equally to this work; names are listed in alphabetical order.}

\begin{abstract}
	Since the early adoption of e-commerce, travel and tourism has been a lab for the design of recommender systems: tools that help travelers choose destinations, flights, accommodations, and combine them into complete itineraries. Data-driven recommendation techniques, ranging from case-based reasoning to reinforcement learning, have been adapted to travelers' needs. The research community has produced multifaceted prototypes of travel and tourism recommender systems (TTRSs), which are context-dependent, multistakeholder-oriented, and more recently, addressing sustainability issues, such as overtourism.
	Despite this enduring work, TTRSs are not widespread yet. We argue that three limitations can explain this fact: outdated and sparse data sets used to train and validate TTRSs, algorithms that prioritize prediction accuracy over domain-specific dimensions such as novelty and contextual relevance, and a failure to address the specific needs of travelers.
	Targeted incremental research could address these limitations, but a disruptive factor has meanwhile entered the ecosystem of tourism information and commercialization platforms: generative artificial intelligence (GenAI). According to market research, simple GenAI applications are becoming the primary entry point for travelers searching and planning their trips. This forces research to rethink how TTRSs should be designed and which core techniques should be integrated. We claim that future TTRSs, in addition to offering personalized information filtering, should become more flexible advisors that support decision making, integrating multiple data types and AI techniques, from data mining to natural language processing. Moreover, they must transparently balance the conflicting goals of travelers, service suppliers, platform owners, and local communities. We then outline research targets for building more effective TTRSs, fruitfully combining old and new recommendation techniques.

\end{abstract}


\maketitle

\section{Travel and Tourism Recommender Systems}\label{sec1}

\subsection{Recommendation tasks}
\Acp{TRS} support travelers and tourists make travel-related decisions, such as which region and points of interest (POIs) to visit, which flights and accommodation to book, and what activities and events to attend. \acp{TRS} can combine multiple types of recommendations and suggest complete travel itineraries the user could follow while visiting a target region~\cite{Ricci2020, DBLP:journals/mta/SarkarMPRP23}. Most \acp{TRS} focus on travelers with tourism motivations, but some of the implemented functionalities, e.g., supporting the selection of an accommodation or a restaurant, are useful for generic travelers. Hence, in this paper, unless we want to stress the specific tourism scenario, for conciseness, we will use the terms travel and traveler.

Throughout the entire travel planning and execution process, \acp{TRS} can support a wide range of travelers' choices. They can suggest: a) tourist destinations, such as cities or regions, b) services, such as hotels, transportation, and restaurants, c) points of interest (POIs) and activities, such as museums, architectural landmarks, concerts, sporting events, and d) complete itineraries, combining multiple types of items arranged sequentially, and compliant with spatial and temporal constraints~\cite{DBLP:journals/tele/Kabassi10, DBLP:journals/jnca/GavalasKMP14, DBLP:journals/jitt/WangHJ25}.

\acp{TRS} can help travelers both before and during their trips~\cite{bellogin2025pointrecommendationpitfallsviable}. In the pre-trip stage, they support the selection of multiple items and services structuring the overall travel plan: the destinations and the POIs to visit, the transportation means, and the accommodation. Then, in the in-trip stage, \acp{TRS} recommend or refine pre-trip suggestions by directing the user to POIs or events relevant to the user's current context, which is only partially known before the travel starts. The information gathered before and during the trip can be used in a post-trip stage to enhance user profiles and improve future recommendations. \autoref{tab:trs_classification} summarizes classic travel recommendation tasks and cites some representative travel platforms that support them. 

\begin{table}[h]
	\centering
	\caption{Broad classification of \Acp{TRS} with respect to the recommendation category. Travel platforms in the last
		column may support more than one recommendation task and be used before and during the trip. In 
		\autoref{s:GenAITourism},
		we describe in more detail some of these and other travel platforms. }
	\label{tab:trs_classification}
	\begin{tabularx}{\textwidth}{p{2.5cm} X p{1.3cm} p{4.5cm}}
		\toprule
		\textbf{Category} &
		\textbf{Typical recommendation task} &
		\textbf{Main stage} &
		\textbf{Representative travel platforms} \\
		\midrule
		Destination recommendation & Recommend cities, regions, countries to visit. & Pre-trip & \href{https://www.google.com/travel/explore}{Google Travel Explore}, \href{https://www.kayak.com/explore/}{Kayak Explore}, \href{https://www.skyscanner.net}{Skyscanner Explore Everywhere}\\
		Tourism service recommendation &
		Recommend tourism products and services, such as
		accommodation, transportation, or other tourism services. & Pre-trip \& in-trip &
		\href{https://www.booking.com/}{Booking.com}, \href{https://www.expedia.com/}{Expedia}, \href{https://layla.ai/}{Layla}, \href{https://mindtrip.ai/}{Mindtrip}\\
		\addlinespace
		POI and/or activity recommendation &
		Recommend individual places or experiences at a destination, including
		attractions, museums, restaurants, landmarks, events, and activities.  & In-trip &
		\href{www.getyourguide.com}{GetYourGuide}, \href{https://www.google.com/maps}{Google Maps},
		\href{https://mindtrip.ai/}{Mindtrip}, 
		\href{https://www.tripadvisor.com/}{Tripadvisor}, \href{https://www.yelp.com}{Yelp}\\
		\addlinespace
		Tour, route, and itinerary recommendation &
		Select and organize multiple POIs or activities into trajectories, routes, or
		multi-day itineraries, often considering spatial, temporal, and user-specific
		constraints.  & In-trip &
		\href{https://layla.ai/}{Layla},
		\href{https://mindtrip.ai/}{Mindtrip},  
		\href{https://roadtrippers.com}{Roadtrippers}, \href{https://www.tripadvisor.com/Trips}{Tripadvisor Trips},
		\href{https://wanderlog.com}{Wanderlog}\\
		\bottomrule
	\end{tabularx}
\end{table}

\subsection{Recommendation Techniques}

\begin{figure}
\centering
\subfloat[\label{fig:trs-interfaces-tradeoffs}]{%
	\includegraphics[width=0.48\textwidth]{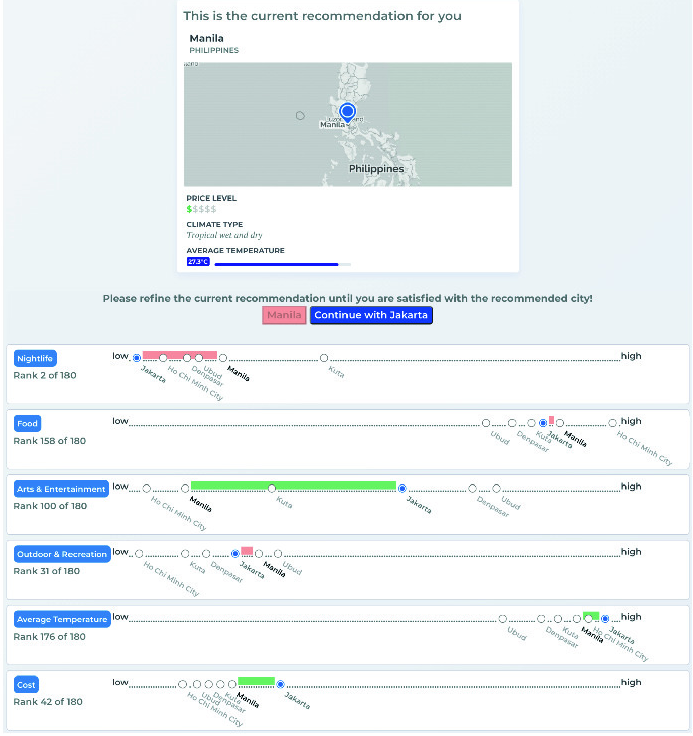}}
\hfill
\subfloat[\label{fig:trs-interfaces-co2}]{%
	\includegraphics[width=0.48\textwidth]{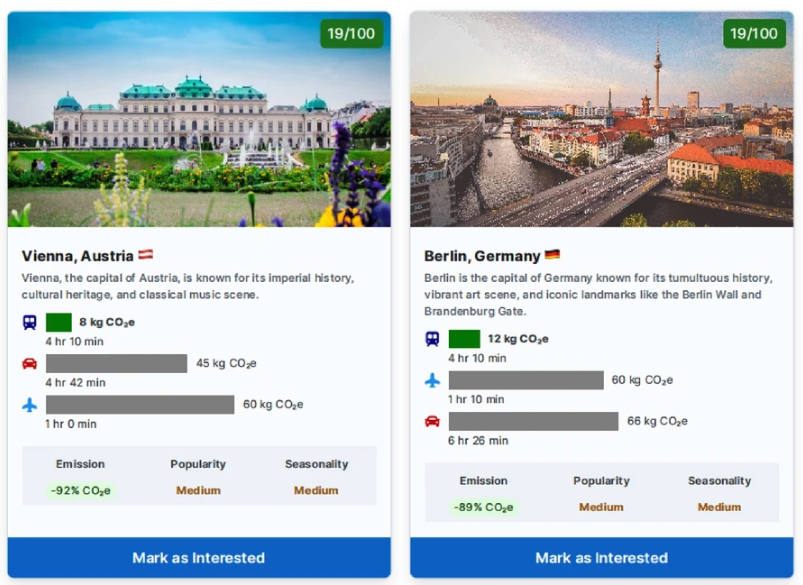}}
\caption{Examples of \ac{TRS} interfaces: (left) refinement of a city recommendation by trading off destination characteristics such as the arts \& entertainment, food, or cost~\cite{Dietz2022ENTER} \textcopyright~2022 Linus W. Dietz, Sameera Thimbiri Palage \& Wolfgang Wörndl; (right) destination suggestions displaying the estimated CO\textsubscript{2} emissions of reaching the destination by train, car, or plane. \cite{Banerjee2025Modeling} \href{https://creativecommons.org/licenses/by/4.0/}{CC-BY} Ashmi Banerjee, Tunar Mahmudov, Emil Adler, Fitri Nur Aisyah \& Wolfgang Wörndl.}
\label{fig:trs-interfaces}
\end{figure}

To support the above-mentioned decision tasks, \acp{TRS} personalize the recommendation process and the interaction with the user. The recommendations may be computed by tailoring them to an individual traveler or to a group of people traveling together~\cite{YeraCM25}. \acp{TRS} can take into account important contextual factors that influence the traveler's decision, such as the location of the traveler, or the forecast weather at the trip time~\cite{Sanchez19,DBLP:journals/ipm/RenjithSJ20}. The user/system interaction design can vary from simple form-based search, driven by manual input of users’ preferences and interests, to more interactive interfaces, based on chat components, 
and web-based conversational interactions~\cite{CamilleriT23,Nguyen2017chat,Dietz2019c}. %
Classical data-driven recommendation techniques, from case-based reasoning to reinforcement learning, have been adapted to the travel scenario~\cite{DBLP:journals/mta/SarkarMPRP23,Massimo023}. We note that the recommendation algorithm must often deal with the new user problem, i.e., the RS rarely has a sufficient amount of historical information about the past user activity to apply sophisticated inferences about what the user may like. Hence, content-based solutions~\cite{MustoGLNS22, DBLP:journals/eswa/BinucciLGLM17}, clustering~\cite{Massimo023, Chang18072022}, and implicit feedback approaches~\cite{JannachLZ18, DBLP:conf/kdd/LianZXSCR14} dominate the field of practical solutions. \autoref{t:recommendationtechniques} lists the most popular recommendation techniques used in the development of \acp{TRS}. 
\autoref{fig:trs-interfaces} shows instead two examples of \acp{TRS} using content-based techniques, for city and points of interest recommendations.

\begin{table}[ht!]
\caption{Commonly used TTRS recommendation techniques. 
}
\label{t:recommendationtechniques}
\begin{tabularx}{\columnwidth}{p{2.15cm}Xr}
	\toprule
	\textbf{Technique} & \textbf{Description} & \textbf{Refs}\\
	\midrule
	Clustering & Group users or items with similar attributes, patterns, or preferences. & \cite{Chang18072022,DBLP:journals/eswa/TliliK21}\\
	Content-based & Suggest items with features matching users' past preferences.& \cite{DBLP:journals/eswa/BinucciLGLM17,Dietz2022ENTER}\\
	Context-aware & Use contextual information (e.g., weather, time, location, season) to adapt to the users' current needs.& \cite{DBLP:journals/eswa/Abbasi-MoudHKE22, Li2015Rank}\\
	Conversational & Collect and refine users' preferences via an interactive dialogue.& \cite{Nguyen2017chat,Dietz2019c} \\
	Deep learning & Use neural networks to model non-linear relationships and leverage heterogeneous data. & \cite{DBLP:journals/eswa/ChenZCWC20,VecchiaMQGB24} \\
	Group recommendation & Produce recommendations for groups of travelers, by combining individuals' preferences. & \cite{DBLP:journals/eswa/AlvesMNNAM24, DBLP:journals/eswa/SarkarM22} \\
	Implicit feedback & Derive user preferences from observed interactions, such as visits or check-ins. 
	& \cite{DBLP:conf/kdd/LianZXSCR14, Li2015Rank}\\
	Metaheuristic optimization & Adapt recommendations under several constraints and objectives using optimization algorithms. & \cite{DBLP:journals/eswa/TliliK21, DBLP:journals/eswa/ChenZCWC20} \\
	Sentiment analysis & Incorporate opinions, attitudes, or preferences extracted from text (social media posts, reviews, etc.).& \cite{DBLP:journals/eswa/Abbasi-MoudVS21, DBLP:conf/bigdataconf/HoLL23}\\
	Simulation & Compare recommendation policies by creating artificial tourism environments and interactions. & \cite{PiliponyteMR24, Merinov024} \\
	Social & Model preferences exploiting social/community interactions between users.& \cite{DBLP:journals/eswa/EsmaeiliMGM20, Chang18072022}\\
	Sustainability & Generate recommendations adding objectives related to environmental sustainability such as social or seasonal factors. & \cite{VecchiaMQGB24,Banerjee2025Modeling} \\
	Trajectory sequential mining & Learn mobility patterns and transitions between POIs to compute route recommendations.& \cite{DBLP:conf/bigdataconf/HoLL23,Dietz2020} \\
	
	\bottomrule
\end{tabularx}
\end{table}

\subsection{Application Domain Related Challenges}

The range of technologies discussed previously is motivated by the system-supported tasks. However, to better understand what technologies should be used and improved, it is fundamental to single out the specific challenges of the application domain. The travel and tourism domain is characterized by distinctive features, which make it quite different from other typical \acp{RS} domains (e.g., music or movies)~\cite{MassimoR22}. 

Firstly, \acp{TRS} should satisfy a wide spectrum of heterogeneous travelers, who travel with diverse motivations and needs. Moreover, the intent of the traveler and the attractiveness of alternative travel targets, as we stressed above, are context-dependent, based on season and mobility constraints, but also on highly variable social trends and items' popularity. Various contextual aspects are linked to social dimensions: traveling is often a group activity (family, friends, etc.), and travelers' choices are influenced by their exposure to social networks. 

Secondly, many of the system-recommended ``items'' (in the standard RS jargon), for instance, a destination like the Loire Valley in France, cannot be objectively modeled by measurable features only. Important properties of tourist items, which must be used in the recommendation process, refer to the holistic experience the traveler may have. The assessment of these properties will challenge the users when they try to anticipate their value while interacting with the system, and evaluating the recommendations. An attraction is not ultimately selected because of its physical characteristics, but because its image and story match the desired emotion and experience of the traveler~\cite{DBLP:journals/expert/StaabWRZGFPK02}.

Thirdly, the RS must produce revenues for the hosting platform within a tightly networked economy, involving a range of diverse stakeholders~\cite{WerthnerACDGJNP15}. The provider's side is composed of large online booking platforms (flights and accommodations) and global travel and hospitality chains. However, small local service providers (e.g., tour guides and rural accommodation chains) are also present and have regional specializations. These stakeholders have conflicting objectives, and their interests diverge from those of the travelers and the local communities living in touristic areas, which may even suffer from the excessive touristic development of their region (overtourism)~\cite{Dodds2019phenomena,Merinov024,Ricci2025TTRS}. All these stakeholders, suppliers, consumers, and local communities, look for specific benefits and therefore have different requirements for what the RS should do for them~\cite{bellogin2025pointrecommendationpitfallsviable,AbdollahpouriB22}. 
As a related aspect, travel planning decisions are high-stakes, as they involve substantial investment of time and money, as well as binding booking commitments. Hence, reliability and trust in the RS are essential~\cite{DBLP:journals/umuai/AlvesMSCNM23, DBLP:journals/expert/StaabWRZGFPK02, DBLP:journals/jitt/WangHJ25,Ricci2025TTRS}. However, travelers' trust in the system is challenged by the commercial nature of the platform, whose revenues are also dependent on commission-based recommendations~\cite{Bapiri23072024, DBLP:journals/fdata/BanerjeeBW24}.

\subsection{Pitfalls of Travel RSs}

Notwithstanding the above-mentioned specificity of the travel and tourism application scenario, a large portion of the research has uncritically followed the main conceptualization of RSs: algorithms predicting travelers' preferences and choices by using recorded logs of travelers' spatiotemporal movements, bookings, and visits. 
Data sets with this information have been mined with general-purpose recommendation techniques (see \autoref{t:recommendationtechniques}). 
However, targeting preference and behavior prediction with standard RS techniques is more difficult to implement in travel and tourism domains because of the discussed complexity of the recommendation scenario. Moreover, logged user-item interactions, which are stored in open data sets, are often much more sparse and outdated, compared to those available in other domains (media). This aspect, combined with the mentioned fact that users are typically new to the system (no recorded interactions), poses a hard technical obstacle for the recommendation algorithm. Moreover, even if the user has already interacted with the RS, their choices are highly context-dependent, making it unlikely that the system has enough data about users' behavior to generate reliable recommendations for a novel contextual condition. This combination of factors results in insufficient data sources and therefore poorly personalized and contextualized recommendations. Thus, in general, recommendations are composed mainly of popular, non-personal suggestions and suffer from poor diversity~\cite{Abdollahpouri2019Popularity,bellogin2025pointrecommendationpitfallsviable}. These pitfalls of \acp{TRS} will be more extensively discussed in the next section of the paper. The concrete result is that, despite the ongoing research activity on \acp{TRS}, their adoption in operational travel and tourism platforms is limited: even major platforms still advertise popular destinations and attractions and use simple preference acquisition interactions and inference methods. 

\subsection{Building Radically Novel Solutions}

Additional and incremental research, along the classical sub-fields of RSs, could face the above-mentioned challenges and limitations. However, a disruptive factor is penetrating the ecosystem of tourism information and commercialization platforms: generative artificial intelligence (GenAI). Rather simple applications based on GenAI tools, which are described in this paper, are rapidly becoming the primary entry points for tourists to search and plan their travels. Recent advances in large language models are producing new types of conversational recommendation and decision support interactions that travelers are increasingly using and relying on. They enable more flexible and interactive forms of information search and recommendation~\cite{10.1145/3626772.3657840,Park2025Enhancing,Palahan2024Innovating}. Nevertheless, they also pose their own issues, related to reliability, coverage, and transparency. 

In this paper, we start by reflecting on the research field of \acp{TRS}, and we identify some reasons why it has produced limited practical impact. We refer to recent extensive analyses discussing a structured collection of pitfalls of \acp{TRS}~\cite{bellogin2025pointrecommendationpitfallsviable,DBLP:journals/is/WerneckSVPMR21} (\autoref{sec:pitfalls}). We believe that new GenAI techniques should encourage the RS research community not only to fix technical issues, but to design radically novel and better solutions for all the involved stakeholders, which address the true challenges of the application domain, which are not only technical but also social and economic. This new generation of \acp{TRS} should be based on the integration of traditional data-driven RS approaches, still not fully exploited by GenAI techniques, and the powerful information processing capabilities offered by large language models. To foster this transition, 
we first review the available GenAI tools that have been deployed and are used today by travelers in a range of application scenarios and tasks (\autoref{s:initialAdoption}). Later, we summarize important findings about users' opinions on \ac{GenAI} applications to search for travel and tourism information (\autoref{s:GenAIUsage}).
Finally, we combine information derived from current \acp{TRS} pitfalls with the analysis of the potential of GenAI solutions, aiming at providing guidance on how the evolving field of \acp{TRS} can address long-standing challenges and better align with user, business, and societal needs (\autoref{sec:disruption_genai}). 

\section{Pitfalls in Travel and Tourism Recommender Research}
\label{sec:pitfalls}

\ac{TRS} research is an established branch of RS research with an active and multidisciplinary community. The RSs in Tourism Workshop has been organized 10 times in conjunction with the ACM RecSys Conference, and each year both venues feature a dozen papers on \acp{TRS}. Other venues, combining scientific and industrial focus on ICT and Tourism, are the annual ENTER conference and the Journal on Information Technologies and Tourism. Both of them have published numerous papers on \acp{TRS}. However, despite this substantial research activity, \acp{TRS} are not yet widely used in practice. Although some start-ups have offered solutions for intelligent trip planning and context-aware in-trip recommendations, before the introduction of generative AI, the commercial applications of \acp{TRS} in the major market platforms are limited to simple forms of information filtering and generic inspirational interfaces. 

In~\cite{bellogin2025pointrecommendationpitfallsviable}, this lack of system adoption is attributed to technological deficiencies in three dimensions: data sets, recommendation algorithms, and system evaluation methods. We summarize these deficiencies by following~\cite{bellogin2025pointrecommendationpitfallsviable,DBLP:journals/is/WerneckSVPMR21,DBLP:journals/jitt/WangHJ25}. \autoref{fig:disruptive-potential} (left) shows the most important identified pitfalls in these dimensions.

\subsection{Data Sets Pitfalls}

A major obstacle to the design and validation of effective \acp{TRS} is the lack of up-to-date travelers' behavior data repositories: publicly available data sets record users' visits to tourism venues observed more than 10 years ago. Nowadays, some venues are no longer open, have changed categories, or have altered the services they offer to users. To our knowledge, no new data sets related to the recommendation of travel and tourism items have been published since the COVID-19 pandemic. 
Researchers still rely on old \acp{LBSN} check-ins from \href{https://foursquare.com/}{Foursquare}, \href{https://www.yelp.com/}{Yelp}, \href{https://snap.stanford.edu/data/loc-Gowalla.html}{Gowalla}, or geolocated pictures (e.g., \href{https://www.flickr.com/}{Flickr}), for their offline analyses of \acp{RS}. Furthermore, these mobility data sets do not originate from travelers' interactions with a \ac{RS}: they collect user actions on a social network platform, signaling only users' implicit preferences. In fact, a location check-in or an image upload only says that the user visited a place, and maybe it was liked. Moreover, the timestamp of the log data is often later than the actual visit, hence it cannot be reliably used to extract visit context information or to sequence the user's visit trajectory.  Even more importantly, the mentioned data sets are sparse~\cite{DBLP:conf/cikm/BrilhanteMNPR13,Dietz2020} and biased~\cite{DBLP:journals/jitt/WangHJ25}, e.g., by the popularity and the self-selection biases. If not properly managed in the model training phase, these data characteristics reduce the validity of the learned travelers' preference model~\cite{DBLP:journals/csur/SanchezB22}. 

It should also be noted that in \ac{LBSN} data sets, users show mostly local behaviors, with check-ins concentrated in their home city, and many venues are labeled as private residences~\cite{DBLP:conf/um/SanchezD22}. This limits their representativeness for modeling real travelers' behavior and requires extensive pre-processing, to isolate different categories of travelers. Pre-processing varies across studies, which limits the comparability of results. Furthermore, neither tourists nor locals should be treated as homogeneous groups~\cite{Dietz2020}. Local residents may exhibit different mobility patterns depending on whether they are involved in leisure, work-related, or family activities, but it is hard to precisely extract this contextual information from the data. Similarly, tourists may differ according to their origin, travel purpose, duration of stay, and other characteristics of the trip, including domestic, international, and business travelers.
In practice, many of these data sets lack key information for a \ac{RS}, such as demographics, textual reviews, and context, such as the travel purpose, the visited locations' crowdedness, or the group composition. These deficiencies limit the usability of the data sets to generate truly personalized recommendations.

In contrast, targeted user studies of \acp{TRS} have collected richer data sets including explicit preference assessments and subjective opinions of venues, but also have their limitations. User studies involve a low number of participants (in general, between 100 and 200), mostly from Western, educated, democratic, and rich populations. They are conducted over a short period of time, and cannot adequately measure important aspects such as seasonal behavior, preference evolution, repeated visits, etc.~\cite{DBLP:journals/csur/ZangerleB23}. Besides, in user studies, participants are instructed to select hypothetical venues in what-if scenarios, without then traveling to the location or spending actual money.

\subsection{Algorithms Pitfalls}

The data set pitfalls mentioned before have strongly influenced the development of recommendation algorithms. The issues of sparse interactions between users and venues, implicit preference information, and new users (users without log data in the RS), have been typically addressed by using models that exploit recommended item descriptions (content) and implicit feedback, e.g., POI visits. In addition to algorithmic issues caused by poor data quality, other limitations come from the heuristics of the recommendation algorithm and the optimization/learning method.

An important algorithmic pitfall is that many traditional \acp{RS} are affected by the high levels of exposure and popularity biases present in available tourism data~\cite{Banerjee2020Analyzing}. We can observe this in models based on implicit feedback that learn user preferences through the check-ins they perform. Since popular POIs have already received many visits from users in the past, implicit feedback models might even reinforce the recommendation of popular locations. Although content-based approaches could offer solutions to address this issue by exploiting the attributes of items, user profiles might still be built on these biased interactions. In \acp{TRS}, this bias analysis is especially relevant, as recommending the most popular venues does not provide relevant information to tourists, who can easily find popular spots using travel guides.

Furthermore, tourism recommendations affect numerous stakeholders with potentially conflicting objectives. Hence, travel recommendation problems should be addressed as multi-objective tasks~\cite{Merinov024}, in which the interests of travelers must be considered along with the heterogeneous needs of other stakeholders such as local communities/governments, destination management organizations (DMOs), service providers, etc. For example, DMOs and local communities might be more interested in aspects related to sustainability and overtourism. E-commerce platforms may be required to meet specific business goals, and some providers may demand that their services receive unbiased visibility/exposure. However, most current recommendation approaches are optimized to satisfy user interests and platform owner revenues, ignoring other perspectives.  

Another discrepancy in model optimization is especially evident when we analyze the contextual and dynamic nature of trips. Tourism choices are strongly influenced by the context of the visit (in-trip), even more than by long-term user preferences for certain types of attractions~\cite{MassimoR22}. Context is not limited to temporal or geographical distances. It also refers to the mood of the user, the weather, current intent, group interests, the phase of the trip, or budget, among other factors. Besides, tourists' decisions are highly dynamic, something that clashes with the rigidity of many algorithms during the in-trip recommendation phase, as they do not account for changes in the tourist's intentions. A tourist who prefers museums may enter a nearby restaurant or cafeteria due to fatigue or may change plans due to a sudden weather change. One possible solution to this problem could be the use of conversational systems, as they allow users to update their interests in real time. However, although there are proposals for interactive conversational systems, they are not the predominant approach in academia~\cite{DBLP:journals/jitt/WangHJ25}. 

Meanwhile, from the user's perspective, visiting a recommended POI tends to require a substantial amount of time, money, and effort. Therefore, the recommendations need to be relevant, but also, \Acp{TRS} should provide reasons why these POIs have been recommended. This is especially important when these recommendations are generated by opaque black-box models that may not be understood by the traveler. Hence, to increase travelers' trust in the system, \Acp{TRS} should provide useful explanations that link recommended POIs to the user's actual context and preferences. This would also allow travelers to determine whether the recommendation fits their profile.

\subsection{Evaluation Methodology Pitfalls}

Evaluation methodologies of \Acp{TRS} are largely based on evaluation approaches used in other traditional recommendation domains. Many \Acp{TRS} are evaluated in an offline scenario, in which the travelers' interactions are divided into at least two disjoint sets: training and test. Then, the recommendations generated using the training set are compared with the visits the users have performed in the test set. Although this is a standard protocol, it does not provide a complete picture of the effectiveness of the system, and its potential in changing the organic behavior (attracted by popular and fashionable choices). Furthermore, current \Acp{TRS} evaluation methodologies are also based on generic ranking accuracy metrics, apply inconsistent testing protocols (limiting the reproducibility of the results obtained), do not consider contextual factors, and group heterogeneous users together~\cite{DBLP:journals/csur/SanchezB22, DBLP:journals/jitt/WangHJ25,DBLP:conf/um/SanchezD22}.

A fundamental methodological pitfall in evaluation is the discrepancy between the metrics used to measure the performance of \Acp{TRS} and the actual goal of \Acp{TRS}, that is, to support travelers throughout the decision-making process, not just to predict their interests. Standard evaluation is modeled in a simplified process by using classical offline metrics to measure how well a recommendation model is able to reproduce the missing interactions of the training set (i.e., it is generally accepted that correctly predicting the items the user has chosen in the past is an indicator that the recommendations are relevant). Hence, the standard evaluation methodology does not measure how the RS addresses other important aspects of the tourism sector: to allow travelers to explore and discover new services or attractions, to support the goals of multiple stakeholders, and to favor complementary goals of the recommendations, such as novelty, diversity, or catalog coverage~\cite{DBLP:journals/jitt/WangHJ25}.

Apart from that, there are substantial differences between different groups of users (locals vs tourists, domestic vs international tourists, cold-start vs warm users, etc.), venues (popular vs niche POIs, POIs grouped by categories, etc.), and evaluation aspects (seasonal periods, different cities, etc.)~\cite{DBLP:journals/csur/SanchezB22}. Hence, an algorithm can improve the observed global performance just because it is more accurate for specific types of users or by recommending a specific type of POI in a particular period, while worsening the experience for other users, venues, and stakeholders. 

At the same time, the performance of the models might vary 
substantially depending on the specific conditions under which they are evaluated. Although studies that remove users and items with very few interactions and use time-dependent splits are common, there does not seem to be a standard evaluation protocol, making it difficult to compare the proposed methods and determine the validity of the reported performance improvements. This is also related to reproducibility, since not all researchers accurately and completely describe the adopted evaluation procedure~\cite{DBLP:journals/csur/SanchezB22}.

Finally, \acp{TRS} are inherently interactive, so their evaluation should also assess factors centered on human/computer interaction, such as system accessibility, usability, usefulness, and complexity, which shape user trust and adoption in practice~\cite{bellogin2025pointrecommendationpitfallsviable}.

\section{GenAI and Tourism Information Search}
\label{s:GenAITourism}

Six research directions have been suggested to overcome the previously mentioned pitfalls, namely: multistakeholder design, context-awareness, unbiased data collection, trustworthy design, novel conversational interactions, and real-world evaluation~\cite{bellogin2025pointrecommendationpitfallsviable}.
However, these research lines, as in \emph{``normal''} scientific research, in the sense of T. Kuhn~\cite{Kuhn2012structure}, move from research gaps to hypotheses about viable solutions. In particular, they are based on the consolidated conceptualization of RSs as technological components, embedded in information or selling platforms, that manage items to be recommended. Such an approach may be neither optimal nor far-sighted. %

In the era of social networks and GenAI bots, travel planning and the whole travel experience are changing~\cite{jung2026genai}. Recent tourism literature has shown that social interactions play a predominant role. People move to replicate experiences and be part of a social community~\cite{doi:10.1177/2158244020925511}. Widely used sources include algorithmically controlled platforms and AI bots, which provide a selected view of the available, possibly biased, and outdated information~\cite{JIA2026105442}. In this section, we address this new scenario.

\subsection{Initial Adoption of GenAI}
\label{s:initialAdoption}

Recent advances in \ac{GenAI} are changing the digital travel ecosystem. Travelers now search for information, state their preferences, compare options, and interact with travel services in new ways, where \ac{GenAI} is no longer confined to stand-alone chatbots.
Established travel platforms are incorporating 
\acp{LLM}-powered features into their search, recommendation, and booking systems, adding context-aware recommendations and iterative trip planning in natural language. \ac{GenAI} tools thus act as intermediaries throughout the travel life-cycle, making interactions more adaptive, personalized, and automated~\cite{jung2026genai,Li2025Generative}.
We now discuss some notable systems using the categorization in~\autoref{tab:trs_classification}.

\paragraph*{Destination recommendation}~
Travel platforms such as \href{https://guidegeek.com/}{GuideGeek}, \href{https://gemini.google.com/}{Google Gemini}, \href{https://kayak.com/}{Kayak}, or \href{https://layla.ai/}{Layla}, allow travelers in the pre-trip stage to describe their interests, budget constraints, and type of trip in natural language, instead of working only with predefined filters. These conversational agents can refine the user's needs during successive interactions and merge them with external travel information. For example, \href{https://guidegeek.com/}{GuideGeek}'s architecture combines \acp{LLM} such as GPT-4 with indexed content from the destination's website. If the internal data is not sufficient to answer the traveler's request, the query is forwarded to a web search. Similarly, \href{https://kayak.com/}{Kayak} uses ChatGPT functionalities to support queries that prioritize cities based on the travelers' interests (e.g., architecture, food, etc.).

\paragraph*{Tourism service recommendation}~
This task is supported in both the pre-trip and in-trip stages by travel platforms like \href{https://guidegeek.com/}{GuideGeek}, \href{https://gemini.google.com/}{Google Gemini}, \href{https://kayak.com/}{Kayak}, \href{https://layla.ai/}{Layla}, \href{https://mindtrip.ai/}{MindTrip}, or \href{https://tripplanner.ai/}{Tripplanner}, as these tools allow travelers to compare transportation, accommodation, excursions, and plans, sometimes with real-time prices obtained from external public information or data APIs. This can help travelers to refine the available options according to their wishes.
\href{https://layla.ai/}{Layla} combines interactive communication with the traveler with real-time information on hotels, trains, flights, etc. Through the conversational chat, travelers can search for any of these services to verify if they fit their budget, ratings, and features.
Their platform is connected to other providers like \href{https://www.booking.com/}{Booking.com}, \href{www.getyourguide.com}{GetYourGuide}, and \href{https://www.skyscanner.net}{Skyscanner} and also uses a prediction engine to forecast flight prices. \href{https://mindtrip.ai/}{MindTrip} provides specific conversational agents that use real-time information for both flights and hotels, helping the user filter options based on the criteria they specify (money, flexible dates, availability, etc.). It also offers support for searching for events and activities. However, reservations still need to be made through third-party providers.

\paragraph*{POI and activity recommendation}~
To solve queries for this recommendation category, GenAI leverages additional contextual information when processing conversations.
Travel platforms like \href{https://airial.travel/}{Airial}, \href{https://copilot2trip.com/}{Copilot2trip}, \href{https://guidegeek.com/}{GuideGeek}, \href{https://gemini.google.com/}{Google Gemini}, or \href{https://layla.ai/}{Layla}, support this type of recommendation.
In this scenario, when travelers ask for attractions like museums, restaurants, or events, the travel platforms can incorporate supplementary information regarding ratings, venue details, schedules, or weather, from internal or external data sources like social networks. In this case, \href{https://copilot2trip.com/}{Copilot2trip} offers POI and activity recommendations using manually selected maps and POIs with AI-generated content. Travelers can ask for attractions, lesser-known venues, tours, or experiences and incorporate these recommendations into an interactive map.
\href{https://gemini.google.com/}{Google Gemini} uses public information from Google Maps to provide details about POIs, such as timetables, directions, and ratings, and can take advantage of the user's current location to find interesting venues nearby. 
\paragraph*{Tour, route, and itinerary recommendation}~
Some platforms like \href{https://airial.travel/}{Airial}, \href{https://copilot2trip.com/}{Copilot2trip}, \href{https://layla.ai/}
{Layla}, \href{https://tripplanner.ai/}{Tripplanner}, and \href{https://wonderplan.ai/}{Wonderplan}, show how GenAI can combine individual recommendations into complete travel itineraries. To do so, some of these systems generate itineraries considering constraints such as transport options, venue schedules, and geographical proximity, among others.
In some cases, these itineraries are flexible, allowing users to modify the recommendations by adding venues or by reducing or increasing the time spent at other locations. \href{https://airial.travel/}{Airial} combines LLM-based interactions with constraint-based optimization and dependency graphs to manage itinerary logistics, considering when activities should take place, as well as the mobility patterns and preferences of travelers. \href{https://layla.ai/}{Layla} proposes daily plans considering availability and money constraints, also adapting to the traveler's current interests. It also supports road trips and itineraries involving different cities.

It becomes clear that current GenAI approaches coexist with the information sources available on travel platforms, external information sources and APIs, and up-to-date external travel information.
This allows travelers to interact with the platform more flexibly, enabling them to communicate with it through language models and thus refine their needs and requirements in a more personalized way.

\subsection{GenAI Tool Usage and Effectiveness}
\label{s:GenAIUsage}

Since 2024, chatbots have appeared on corporate websites with little evident link to user needs.
Such \emph{``AI washing''} suggests that firms feel pressed to signal AI adoption, often with little measurable return~\cite{MITNanda2025} and even at the cost of customer trust~\cite{Song2026AI}.
This leaves many firms stuck between experimentation and integration of novel GenAI solutions~\cite{Balazs2026}.
This failure on the supply side does not mean demand is lacking. The systems surveyed in \autoref{s:initialAdoption} address concrete travel tasks, and, according to surveys, travelers are adopting GenAI quickly and find it useful, even though trust remains limited.

Recent consumer surveys show a rapidly evolving landscape for GenAI applications, revealing new and emerging usage scenarios.
Phocuswright, an independent market research agency, has found that more than half of U.S.\,travelers have tried GenAI in some form, with one in three already using it for travel-related purposes. Moreover, 78\% of travelers found the GenAI results to be somewhat or very helpful for both planning and destination use. However, only about one-third of travelers say that they fully trust GenAI responses~\cite{Schmid2025Chat}. %

In Japan, JTB Tourism Research \& Consulting Co.\,has found that 77.8\% of GenAI users have used it for travel-related purposes---especially younger users, with 89.3\% of women in their 20s and 85.7\% of men in their 30s having used AI for travel.
When asked about the future use of GenAI, top responses included ``creating travel itineraries and suggesting routes'', ``searching and booking transportation options'', and ``searching for local gourmet information''. Moreover, ``explaining tourist attractions'' and ``predicting congestion levels'' were indicated as functionalities where users expect helpful real-time information. It is worth noting that GenAI is perceived not only as an information processing tool, but also as a ``trusted advisor'', ``reliable friend'', and ``partner''~\cite{TBTourism2025}.

However, big market players are still evaluating the prospect of introducing GenAI. Booking.com has conducted its own survey and has found that 91\% of the respondents say that they are excited about AI and 79\% are familiar with the technology, though levels vary between regions. However, despite widespread enthusiasm, only 6\% fully trust AI and the majority (91\%) have at least one concern about its implications. Only 12\% of consumers are comfortable with AI making decisions independently, but still 89\% of consumers want to use AI in future travel planning, with AI assistants (24\%) now considered a more trusted source than travel bloggers (19\%) or social media influencers (14\%)~\cite{Booking.com2025Global}.

Various surveys and categorizations of \ac{GenAI} use have also appeared in the scientific literature~\cite{ilieva2024effects,jung2026genai,zhang2026effects}. Researchers identify the features, benefits, and expectations of using \ac{GenAI} in different tourism settings, some of them overlapping with the functionalities presented before but also covering image or video generation~\cite{zhu2025tourism}, virtual assistants~\cite{loureiro2022role}, and voice interfaces~\cite{firmansyah2023application}. There are also efforts to measure the quality of these tools in tourism~\cite{ilieva2024effects}. The main conclusion is consistent with earlier findings: they add value to the travel experience, have the potential to significantly transform the tourist experience, but trust and privacy issues are commonly raised~\cite{kang2026can}, confirming that these tools can be used as starting points, rather than final ones, in the travel planning process.
Critical analyses have also examined their application to review summaries~\cite{corbin2025missing}, trip planning~\cite{kang2026can}, or the effects on consumer behavior~\cite{zhang2026effects}.
In the next section, we summarize the open issues and propose some lines of action.

\section{The Disruptive Potential of GenAI}
\label{sec:disruption_genai}

\begin{figure}[t]
\centering
\includegraphics[width=\textwidth]{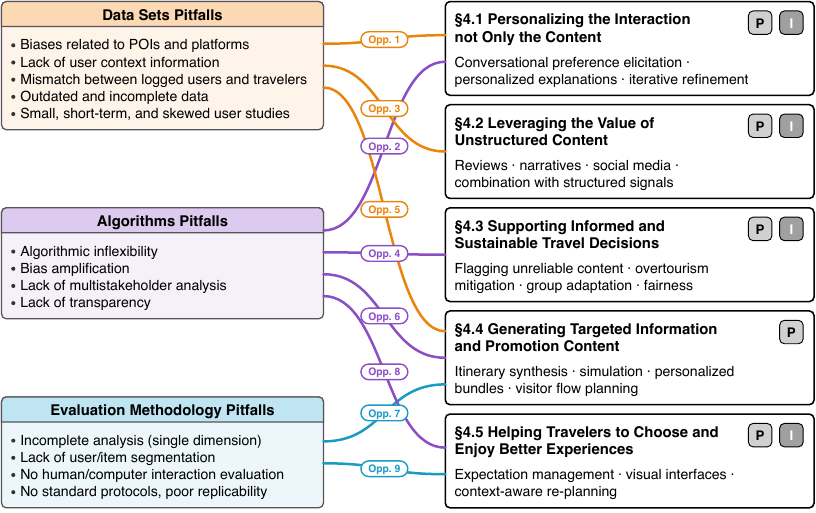}
\caption{GenAI provides great potential to overcome current pitfalls in \ac{TRS} (left). We identified five research goals for integrating \ac{TRS} techniques with \ac{GenAI} (right). The identified connections between pitfalls and research goals are opportunities (Opp.~1--9).  Symbols \PreTripBadge~and \InTripBadge~represent the pre-trip and in-trip phases of recommendation, respectively.}
\label{fig:disruptive-potential}
\end{figure}

The preceding analysis reveals a general change in how travelers discover and select tourism services. Rather than relying on a single \ac{TRS} that profiles users, filters suggestions of candidate items, and presents a ranked list of them, travelers now query \ac{GenAI} tools, cross-reference results with traditional sources, and ultimately make their decisions~\cite{Li2025Generative}.
Filtering and ranking algorithms are still important but are absorbed into conversational, multi-source information flows, where the traveler better controls the process~\cite{DBLP:journals/tors/Deldjoo26}.
Furthermore, this novel approach renders information filtering systems anachronistic as standalone solutions, implying that RSs will move \emph{``under the hood''} of agent-powered natural language models.

However, current GenAI-based tools still face significant limitations, mainly related to their reliability and coverage, which is slowing broader industry adoption. Current systems often produce factual inaccuracies, e.g., incorrect walking times or stale event listings, and they offer no mechanism for learning user profiles from individual user feedback within a session or across trips. Most importantly, their context cannot effectively leverage thousands of items and many more logs of users' preferences. These are precisely the challenges that classical \ac{RS} research has addressed through preference modeling, structured metadata exploitation, and feedback loops, e.g., in conversational refinement of preferences~\cite{Jannach2021,Dietz2022ENTER}.
Discarding this body of work in favor of RSs built purely with general-purpose language models would mean re-encountering solved problems without the evaluation rigor that the \ac{RS} community has developed over two decades~\cite{JannachA16,bellogin2025pointrecommendationpitfallsviable,Deldjoo2023Fairness}.

We therefore argue that the path forward is not \ac{GenAI} replacing \acp{TRS}, but a convergence~\cite{DBLP:journals/jitt/FriedliDS25}: recommendation obtained within a collaborative multiparty advisory process that combines \ac{GenAI}'s conversational flexibility and content-generation capabilities with the strengths of \acp{RS} in preference learning, structured-data exploitation, and evaluation methodology.
In this context, \ac{RS} solutions should shift from rigid information filtering tools to more flexible and dynamic conversational advisors that can leverage the capabilities of LLMs to create more adaptable and interpretable services serving all the stakeholders of the travel and tourism market~\cite{AkataBRDDEFGHHH20}.
In this perspective, we have identified five research goals that could leverage the integration of traditional RS capabilities with GenAI technologies. We discuss which of the \ac{TRS} pitfalls each goal can address, and summarize these links as opportunities (Opp.~1--9) in \autoref{fig:disruptive-potential}.

\subsection{Personalizing the Interaction, Not Only the Content}
A clear advantage of GenAI approaches to recommendation is their suitability for implementing conversational solutions~\cite{DeldjooSPB0CM25}, reducing the burden of retrieving useful information and more easily acquiring users' preferences and travel constraints.
These solutions help travelers simplify their travel planning process, making it faster, reducing repetitive tasks, and allowing easier integration of alternative sources~\cite{VolchekI24}.
They also directly target the data set pitfall of poor user context (Opp.~1): rather than inferring a preference model from sparse and outdated interaction logs, the system can elicit travel purpose, group composition, budget, or mood in natural language at the moment the recommendation is needed, and feed this context into the preference model of the \ac{RS}.
However, what is still needed are techniques that can better personalize \emph{how} the RS interacts with the user, i.e., how the interaction unfolds and iterates the full range of communication steps, including preference elicitation, recommendation, and refinement.
The goal is to design recommendation interactions that can produce more effective %
explanations, by personalizing the interaction itself~\cite{Sun2026When,zhang2026effects,Park2025Enhancing}.
Such novel user interaction approaches can address two of the algorithmic pitfalls (Opp.~2). Firstly, the rigidity of one-shot, top-$k$ recommendation, since an iterative dialogue lets the traveler reveal a change of intent, e.g., fatigue or a sudden change of weather, and the system re-plan accordingly. Secondly, the lack of transparency, since natural language explanations can make the reasons behind a suggestion inspectable, provided they faithfully reflect the underlying selection process rather than merely rationalizing it.

\subsection{Leveraging the Value of Unstructured Content}
Travel and tourism platforms hold vast amounts of reviews, narratives, and social media content that traditional \acp{RS} have largely overlooked, mainly because extracting reliable preference signals from such unstructured data is highly complex. GenAI tools can now effectively process this content, which has proven to be more persuasive and informative for travelers than structured data alone~\cite{Sun2026When,zhang2026effects,DBLP:journals/jitt/CossatinMFA25, DBLP:conf/recsys/WangHJ25}.
Unlocking these sources also mitigates the data set pitfalls of outdated and incomplete behavioral logs (Opp.~3): reviews and narratives are continuously written and carry rich signals about the experiential qualities of venues that check-in data cannot capture, so an \ac{RS} that exploits them no longer depends solely on decade-old interaction logs.
However, this content becomes useful for recommendation only when combined with structured signals such as ratings, timetables, and contextual data~\cite{DeldjooSPB0CM25}.
Rather than serving as opaque drop-in replacements for classical \acp{RS} algorithms, GenAI solutions should enable fundamentally new multimedia recommendation approaches, as early work in fashion recommendation has proposed~\cite{DeldjooNRMPBN24}.

\subsection{Supporting Informed and Sustainable Travel Decisions}
Considering the strong emotional content of travel and tourism items, we see the need to introduce radically new roles for \acp{RS} to act as advisors, also capable of coaching travelers to become more rational decision-makers, who can look beyond their very own needs~\cite{ChangEY25,Dudy2025Unequal}.
Such advisors can help in three ways: flag unreliable content, e.g., inflated claims about influencer-promoted \acp{POI}, so travelers can better judge the information they gather; steer choices towards sustainable options, countering overtourism of popular attractions~\cite{VecchiaMQGB24,Werthner22}; and support fairer group travel decisions through more cooperative planning~\cite{RicciD25}.
The second point confronts the popularity bias pitfall of algorithms (Opp.~4), which GenAI alone does not resolve but rather risks amplifying: \acp{LLM} are trained on web corpora that overrepresent famous attractions, so countering this bias requires the deliberate integration of the debiasing and diversification techniques developed by the \ac{RS} community~\cite{Banerjee2020Analyzing}.
Through such guidance, novel RSs can also help travelers better understand the pros and cons of their travel decisions, guiding them towards more sustainable and informed choices that benefit all stakeholders.

\subsection{Generating Targeted Information and Promotion Content} %
Destinations and service providers can also benefit from GenAI-powered recommendation tools, which address the algorithmic pitfall of disregarding the multistakeholder setting (Opp.~6): most current \acp{TRS} are optimized for the tourist only, leaving the other stakeholders of \autoref{sec1} without support, whereas GenAI can generate content and items tailored to their goals.
GenAI can help these stakeholders create new recommendable items (e.g., themed itineraries) or repurpose available content to promote regional tourism development~\cite{WangW23}.
GenAI-based traveler simulation can generate synthetic interaction data where real data collection is infeasible or ethically problematic, easing the pitfalls of outdated data sets and small-scale user studies (Opp.~5)~\cite{DBLP:journals/fdata/BanerjeeBW24}. Recently introduced simulation techniques can, in addition, estimate the effectiveness of such items and strategies before deployment, addressing some evaluation pitfalls. They can complement offline accuracy metrics with an analysis of the outcomes for all stakeholders, e.g., visitor flows and revenue distribution (Opp.~7)~\cite{PiliponyteMR24,Merinov024}.
These RSs can run client-side, on the stakeholders' desk, supporting strategic reasoning that weighs the interests of all parties, e.g., through a conscious distribution of travelers onto different areas of a destination~\cite{DBLP:journals/fdata/BanerjeeBW24}.
Such approaches can improve the agency of local stakeholders and open vast fields of opportunities to combine items into personalized bundles~\cite{Herzog2019} for prospective travelers to improve fairness considerations~\cite{DBLP:journals/tors/Deldjoo26,Banerjee2020Analyzing,Deldjoo2023Fairness,DBLP:journals/fdata/BanerjeeBW24}.

\subsection{Helping Travelers to Choose and Enjoy Better Experiences}
Tourism is an experiential economy~\cite{WerthnerACDGJNP15}, and the experience is shaped by individual decisions, which in turn are influenced by RS technologies both before and during the trip~\cite{Ricci2020}.
Interacting with \acp{RS} can lead to developing wrong expectations about the selected tourism services~\cite{HofschenM023}, and it has been shown that travelers struggle to accurately anticipate the tradeoffs of choosing novel items such as an unpopular POI~\cite{MassimoR22}.
\ac{GenAI} can help travelers estimate the true satisfaction they will gain from visiting a destination or a \ac{POI} and, just as important, counteract its own persuasive pull when it distorts those expectations (Opp.~8)~\cite{zhang2026effects}.
This reframing also speaks to the evaluation methodology pitfalls (Opp.~9): the measure of success is whether the system helps travelers make better choices and enjoy better experiences, not whether it accurately predicts logged check-ins, and future evaluation protocols should reflect this.
We also believe that new visual interfaces can be developed to better enjoy and follow the recommendations when at the destination, such as detailed personalized instructions about how to best visit certain attractions and dynamic context-aware re-planning of the recommended itinerary.

\section{Conclusions}
\label{s:conclusions}
Travelers are adopting \ac{GenAI} tools quickly and find them useful for planning and during the trip, even though only a minority fully trust their answers.
At the same time, we find that \ac{TRS} research is somewhat held back by the field's pitfalls, limiting the practical adoption of \acp{TRS}.
We therefore posit that \ac{TRS} research needs to embrace human-AI collaboration~\cite{AkataBRDDEFGHHH20}, transforming \acp{TRS} into more transparent conversational advisors~\cite{JannachA16}, mitigating the role and effects of opaque algorithms that limit users' trust~\cite{Ricci2025TTRS,Park2025Enhancing}.
Since such advisors serve travelers, destinations, and service providers alike, multistakeholder \ac{TRS} design and evaluation will be key. Interaction design needs to be followed up with user studies, using better protocols that measure evolving engagement, satisfaction, and trust, as well as the novel set of biases arising from \ac{GenAI}~\cite{Dudy2025Unequal}.
The persuasive nudges of such advisors must be monitored for their real-world impact and unintended consequences~\cite{VecchiaMQGB24,Werthner22}, leading to more \ac{GenAI}-powered evaluation and simulation studies that aim to better match traveler decision-making processes and marketing interventions before deployment~\cite{PiliponyteMR24}.
We argue that the technological foundations of recommender systems, built over two decades of progress~\cite{JannachA16,Dacrema2019Are,bellogin2025pointrecommendationpitfallsviable}, should not be set aside.
The future of recommender systems lies in the opportunities that \ac{GenAI} opens for the field: integrating \ac{GenAI} with the methodology of recommender systems can address long-standing pitfalls, %
and it is this integration that we as a community need to put on our research agenda.

\bibliography{wileyNJD-Chicago}

\end{document}